\documentclass[12pt]{emulateapj}

\newcommand{\etal}{et al.~}
\newcommand{\jybeam}{Jy beam$^{-1}$}
\newcommand{\mjybeam}{mJy beam$^{-1}$}
\newcommand{\nthp}{N$_2$H$^+$ $J=1-0$}
\newcommand{\co}{$^{12}$CO $J=1-0$}
\newcommand{\tco}{$^{13}$CO $J=1-0$}
\newcommand{\cs}{CS $J=2-1$}
\newcommand{\kms}{km s$^{-1}$}
\newcommand{\lkha}{LkH$\alpha$}

\shorttitle{Outflows from Herbig Ae/Be Stars}
\shortauthors{Matthews et al.}
\begin{document}

\title{The Molecular Gas Environment around Two Herbig Ae/Be Stars:
  Resolving the Outflows of LkH$\alpha$ 198 and LkH$\alpha$ 225S}

\author{Brenda C.\ Matthews\altaffilmark{1,2},
James R.~Graham\altaffilmark{2},
Marshall D.~Perrin\altaffilmark{2},
Paul Kalas\altaffilmark{2}}

\altaffiltext{1}{Herzberg Institute of Astrophysics, National Research Council
  of Canada, 5071 West Saanich Road, Victoria, BC, V9E 2E7 Canada}
\altaffiltext{2}{Department of Astronomy, University of California,
  Berkeley, CA, 94720-3411, U.S.A.}

\email{brenda.matthews@nrc-cnrc.gc.ca}

\begin{abstract}
Observations of outflows associated with pre-main-sequence stars
reveal details about morphology, binarity and evolutionary states of
young stellar objects.  We present molecular line data from the
Berkeley-Illinois-Maryland Association array and Five Colleges Radio
Astronomical Observatory toward the regions containing the Herbig
Ae/Be stars \lkha\ 198 and \lkha\ 225S.  Single dish observations of
\co, \tco, \nthp\ and \cs\ were made over a field of $4.3$\arcmin\
$\times 4.3$\arcmin\ for each species.  \co\ data from FCRAO were
combined with high resolution BIMA array data to achieve a
naturally-weighted synthesized beam of 6.75\arcsec\ $\times$
5.5\arcsec\ toward \lkha\ 198 and 5.7\arcsec\ $\times$ 3.95\arcsec\
toward \lkha\ 225S, representing resolution improvements of factors of
approximately 10 and 5 over existing data.  By using uniform
weighting, we achieved another factor of two improvement.  The outflow
around \lkha\ 198 resolves into at least four outflows, none of which
are centered on \lkha\ 198-IR, but even at our resolution, we cannot
exclude the possibility of an outflow associated with this source. In
the \lkha\ 225S region, we find evidence for two outflows associated
with \lkha\ 225S itself and a third outflow is likely driven by this
source. Identification of the driving sources is still
resolution-limited and is also complicated by the presence of three
clouds along the line of sight toward the Cygnus molecular cloud.
\tco\ is present in the environments of both stars along with cold,
dense gas as traced by \cs\ and (in \lkha\ 225S) \nthp.
No 2.6 mm continuum is detected in either region in relatively shallow
maps compared to existing continuum observations.
\end{abstract}


\section{Introduction}

In addition to evidence for mass infall (accretion), many
pre-main-sequence stars also possess prominent outflows and
jets. These jets are believed to play a crucial role in removing
angular momentum from circumstellar disks, allowing accretion to
continue \citep{sta04}.  Outflows can extend many parsecs in length,
and can serve as ``fossil records'' of the accretion process,
providing indications of temporal variability of accretion or
precession of outflow axes over thousands of years \citep{mcg04}.
Furthermore, outflows inject turbulence into molecular clouds and may
regulate overall star formation efficiency \citep{mat00,chr05}.

Outflows from young stars may be observed both through shock-excited
emission lines such as [SII] 6716, 6731 \AA\
\citep[e.g.,][]{fin85,con98} and [FeII] 1.257, 1.644 \micron\
\citep{gra87,pyo03}, and also through millimeter rotational lines of
CO and other molecules \citep[e.g.,][]{sch90,arc06}.  Because star
formation is a complex and turbulent process, high angular resolution
observations (i.e., submillimeter or millimeter interferometry) are
usually required in order to disentangle the various motions of
molecular gas around young stars (infall, outflow, rotation, or
turbulence).  Such observations have been used to probe the kinematics
of outflows from a wide range of YSOs, including T Tauri stars such as
HH 30 \citep{pet06} and also deeply embedded Class I protostars such
as in AFGL 5142 \citep{zha07}, which is a protocluster displaying
at least three molecular outflows originating from at least half a
dozen embedded massive protostars.

\begin{deluxetable*}{llrcrc}
\tablecolumns{6} 
\tablewidth{0pc} 
\tablecaption{FCRAO Observational Summary}
\tablehead{\colhead{Source} & \colhead{Species/Transition} &
  \colhead{Frequency} & \colhead{Antenna} & \colhead{Resolution} &
  \colhead{Sensitivity per Channel} \\
& & \colhead{[GHz]} & \colhead{Efficiency} & \colhead{[\kms]} & \colhead{[K]}}
\startdata 
\lkha\ 198 & \co & 115.21720 & 0.45 & 0.06349 & 0.35 \\
 & \tco & 110.20135 & 0.49 & 0.06642 & 0.15 \\
 & \nthp & 93.17378 & 0.52 & 0.07855 & 0.09  \\
 & \cs & 97.98095 & 0.52 & 0.07470 & 0.09 \\
\hline
\lkha\ 225S & \co & 115.21720 & 0.45 & 0.06350 & 0.40 \\
 & \tco & 110.20135 & 0.49 & 0.06642 & 0.16 \\
 & \nthp & 93.17378 & 0.52 & 0.07470 & 0.09 \\
 & \cs & 97.98095 & 0.52 & 0.07856 & 0.09 \\
\enddata 
\label{FCRAOobs}
\end{deluxetable*}

In this paper we consider molecular outflows in an intermediate case
between these two extremes, by investigating outflows originating from
groups of Herbig Ae/Be (HAeBe) stars.  The HAeBe stars are
pre-main-sequence stars of intermediate-mass, i.e., $1.5 \le M/M_\odot
\le 10$ \citep{her60,wat98}, which exhibit infrared and millimeter
excesses indicative of circumstellar dust. While the majority of
outflows studied to date originate from T Tauri stars, many Herbig
Ae/Be stars are known to launch outflows as well \citep{mun94}, in
some cases extending several parsecs in length \citep{mcg04a}.  HAeBe
stars are often found in groups \citep{tes98,tes97}, making high
angular resolution observations critical for resolving the outflows
associated with individual stars.  These intermediate mass stars are
also rarer than later type T Tauri stars due to the steepness of the
Initial Mass Function, which means they are, on average, found at
greater distances than their lower mass counterparts.  Therefore,
higher angular resolutions are required to resolve their outflows
compared to closer, lower mass T Tauris. 

We have searched for wide field and high resolution emission from \co\
in the environments around the HAeBe stars \lkha\ 198-IR and \lkha\
225S based on intriguing results from near-IR AO imaging polarimetry
by \citet{per04}. In addition, we have obtained single dish spectra of
\tco, \cs\ and \nthp\ toward both sources.  We present here millimeter
observations of the \lkha\ 198 and \lkha\ 225S stars and their nearby
group members.  Our high-angular resolution observations resolve
multiple outflows in both of these regions.  The observations and data
reduction techniques are described in $\S$ \ref{obs}.  Results for
\lkha\ 198 and \lkha\ 225S and their surrounding regions are presented
in $\S$ \ref{lkha198} and $\S$ \ref{lkha225S}, respectively.  Our
findings are summarized in $\S$ \ref{sum}.

\section{Observations and Data Reduction}
\label{obs}

\subsection{FCRAO Data}

We obtained \co, \tco, \cs\ and \nthp\ data from the 14~m Five
Colleges Radio Astronomical Observatory (FCRAO) using the array
SEQUOIA on 2004 November 10.  The telescope was pointed toward \lkha\
198 at $\alpha_{J2000} = 00^{\rm h}11^{\rm m}$25\fs97, $\delta_{J2000}
= +59$\degr49\arcmin29\farcs1 and \lkha\ 225S at $\alpha_{J2000} =
20^{\rm h}20^{\rm m}$30\fs65, $\delta_{J2000} =
+41$\degr21\arcmin25\farcs50.  A fully-sampled image covering a field
of view 4.3\arcmin\ $\times$ 4.3\arcmin\ was made of four spectral
line transitions.  The transitions \co\ and \tco\ were observed
simultaneously; similarly, \cs\ and \nthp\ were obtained in a single
observation. A summary of the FCRAO observations is shown in Table
\ref{FCRAOobs}. Each observation was 24 minutes in duration.
Calibrated T$_A^*$ data are delivered from the telescope.  We
converted these data to main beam temperatures by division of the
appropriate antenna efficiency (see Table \ref{FCRAOobs}).  The
conversion to \jybeam\ was done by deriving the conversion factor
based on the observing wavelength and telescope beam:
\begin{equation}
\frac{{\rm Jy}}{{\rm K}} = \left (13.6 \
\frac{\lambda[{\rm mm}]^2}{\theta^{\prime\prime}_{min} \theta^{\prime\prime}_{max}}\right)^{-1}.
\label{convert}
\end{equation}

\noindent where $\theta^{\prime\prime}_{min}$ and
$\theta^{\prime\prime}_{max}$ are the minor and major axes of the
telescope beam. For single dish millimeter telescopes, the axes are
equal and equivalent to the diffraction limit of the antenna.  For
FCRAO, the conversion factor is 23.6 Jy/K.  The beamsize (resolution)
is 46.7\arcsec\ for \co, 48.9\arcsec\ for \tco, 55.0\arcsec\ for \cs,
and 57.5\arcsec\ for \nthp.  The data were taken in the 25 MHz mode.
The rms sensitivity per channel is shown in Table \ref{FCRAOobs} for
each transition.

\begin{deluxetable*}{lrccc}
\tablecolumns{5} 
\tablewidth{0pc} 
\tablecaption{Summary of \co\ Observations at 115.21720 GHz}
\tablehead{\colhead{Source} & \colhead{Date} & 
\colhead{Array} & \colhead{Resolution} & \colhead{Sensitivity per Channel}  \\
& & \colhead{Configuration} & \colhead{[\kms]} &
\colhead{[\jybeam]} }
\startdata 
\lkha\ 198 & 1 March 2004 & B & 0.127  & 4.6 \\
& 4 March 2004 & B & 0.127  & 5.7 \\
& 18 March 2004 & B & 0.127  & 6.6 \\
& 29 March 2004 & B & 0.127  & 3.4 \\
& 26 April 2004 & C & 0.127  & 1.9  \\
& 1 June 2004 & D & 0.127  & 1.8 \\
\hline
\lkha\ 225S & 28 October 2003 & C & 0.127 & 2.7 \\
& 4 March 2004 & B & 0.254 & 0.85 \\
& 21 March 2004 & B & 0.254 & 1.6 \\
& 31 May 2004 & D & 0.127 & 5.9 \\
\enddata 
\label{BIMAobs}
\end{deluxetable*}

\begin{deluxetable*}{lclcc}[b!]
\tablecolumns{5} 
\tablewidth{0pc} 
\tablecaption{Final Parameters of Combined FCRAO + BIMA \co\ Maps}
\tablehead{\colhead{Source} & \colhead{Channel Width} & 
\colhead{Final Beam} & \colhead{Weighting}  & \colhead{rms per channel} \\
& \colhead{[\kms]} & & & \colhead{[\jybeam]} }
\startdata 
\lkha\ 198 & 0.127 & 6.75\arcsec\ $\times$ 5.47\arcsec\ at $-87^\circ$
& natural & 0.66 \\
& 0.127 & 3.39\arcsec\ $\times$ 2.37\arcsec\ at $-82.5^\circ$ & 
uniform & 1.4 \\
\lkha\ 225S & 0.254 & 5.67\arcsec\ $\times$ 3.95\arcsec\ at
$-79^\circ$ & natural & 0.45 \\
& 0.254 & 2.87\arcsec\ $\times$ 2.29\arcsec\ at $-89^\circ$ & 
uniform & 0.70 \\
\enddata
\label{CombinedSummary}
\end{deluxetable*}

\subsection{BIMA Interferometric Data}

We observed the regions surrounding the two Herbig Ae/Be multiple
systems in single pointings with the Berkeley-Illinois-Maryland
Association (BIMA) interferometer\footnote{The BIMA Array was operated
by the Berkeley-Illinois-Maryland Association under funding from the
National Science Foundation.}  \citep{wel96} in Hat Creek, CA. The
data were taken in 2003 October (1 track each) followed by multiple
observations over the period of 2004 March to 2004 June (see Table
\ref{BIMAobs}).  For both regions, we used the same target coordinates
as for the FCRAO observations.

Three configurations of the ten 6.1 m antennas were used to observe
the \co\ line and continuum at 2.7 mm.  An A-array track and a D-array
track toward \lkha\ 225 were not used due to poor phase correlations.
Two B-array tracks toward \lkha\ 198 were rejected for the same
reason.  The D-array, C-array and B-array data have projected
baselines between 2-13 k$\lambda$, 2-33 k$\lambda$ and 3-74
k$\lambda$, respectively.  Table \ref{BIMAobs} contains the
sensitivities achieved per useable track.

The \co\ line was observed utilizing the digital correlator to record
the line in four bands, configured in redundant bandwidths of 12.5 MHz
(256 channels) and 50 MHz (128 channels), giving spectral resolutions
of 0.127 \kms\ and 1.016 \kms\ for \lkha\ 198.  The same configuration
was used for the C-array track of \lkha\ 225S. For the 2004 tracks
toward \lkha\ 225S, we allocated 256 channels each across bandwidths
of 12.5 MHz and 25 MHz, resulting in velocity resolutions of 0.127
\kms\ and 0.254 \kms, respectively.  \lkha\ 198 was also observed in
some \lkha\ 225S tracks, when the latter source was too high in
elevation to be observed.  The lower sideband was used to measure the
continuum emission at 2.6 mm in bandwidths (125 and 75 MHz for the
respective configurations) limited by the spectral line correlator
configuration.  The typical BIMA array continuum bandwidth was 800
MHz.

Phase and amplitude variations were calibrated by observing quasars at
small angular separations from the source fields: 2015+372 and
0102+584 (for \lkha\ 225S and \lkha\ 198 respectively) approximately
every 30 minutes.  The adopted fluxes of these quasars were epoch
dependent and measured against observations of the planet Uranus or
the quasar 3C454.3 when possible.  The calibration was performed using
the MIRIAD (Multichannel Image Reconstruction, Image Analysis and
Display; \cite{sau95}) task MSELFCAL.  Absolute flux calibration was
done using Uranus when observed or by using the derived fluxes of gain
calibrators during the same epoch as our observations (utilizing the
catalogue of fluxes at ``plot$\_$swflux'' on the BIMA
website\footnote{http://bima.astro.umd.edu}).

Subsequent processing of the data, including the combination of data
from different configurations, was done with MIRIAD.  Images were
produced using MIRIAD's INVERT algorithm with natural weighting and
then by variation of the ``robust'' parameter to produce a
uniform-weighted images optimized for both signal-to-noise and spatial
resolution.  The images were then deconvolved using CLEAN and then
convolved with the clean beam using RESTOR.  The resulting resolutions
and sensitivities of the BIMA data are shown in Table \ref{BIMAobs}.
The resulting noise levels in naturally-weighted images are 0.9
\jybeam\ in 0.127 \kms\ channels for the \co\ line emission from
\lkha\ 198 and 0.75 \jybeam\ in 0.254 \kms\ channels for \lkha\ 225S.
The rms levels in the continuum maps are 18.5 and 17.0 \mjybeam\
toward \lkha\ 198 and \lkha\ 225S respectively.  We achieved
naturally-weighted FWHM spatial resolutions of 7.4\arcsec\ $\times$
6.0\arcsec\ and 5.5\arcsec\ $\times$ 4.1\arcsec\ for the continuum
maps of the two fields.

\subsection{Combination of FCRAO and BIMA data}

The \co\ data from FCRAO and the BIMA array were combined to produce
total power data cubes.  The \co\ emission is strongly detected toward
both sources, so we have used the highest possible spectral resolution
data for combination with the FCRAO data.  In the observation of
\lkha\ 225S, one spectral window exhibited evidence of corrupt
spectra; therefore, we utilized a redundant observation in a second
spectral window for that source.  The FCRAO data were reordered to
match the BIMA axes (x,y,v) and rescaled to Janskys. The MIRIAD task
REGRID was then used to interpolate the FCRAO data onto the BIMA data
cube.  We used the linear combination algorithm of \citet{sta99} to
combine the naturally-weighted BIMA array dirty map with the single
dish map, using the ratio of the beams as the weighting factor.  The
combined image was then deconvolved and convolved with the clean beam
to reconstruct the total \co\ emission.  Maps were created with
natural and uniform weighting. The resultant resolutions and
sensitivities are summarized in Table \ref{CombinedSummary}.  The
primary beam of the BIMA array antennas at 115 GHz is 1.8\arcmin,
which is the area of the resultant combined data maps.

\begin{figure*}
\vspace*{10cm}
\includegraphics{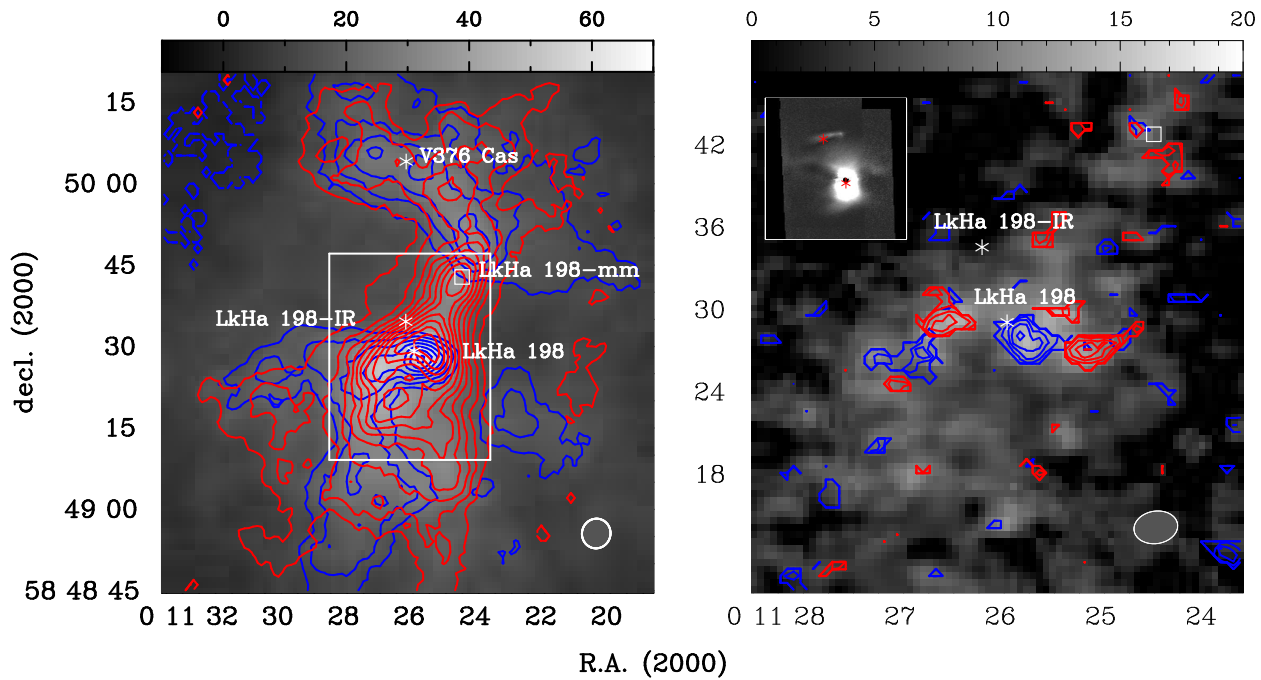}
\caption{{\it Left:} The greyscale shows the zeroth moment over the
entire range of CO emission ($-4$ to 4 \kms) of the combined FCRAO +
BIMA data toward the region surrounding \lkha\ 198 (in units of
\jybeam\ \kms). Optically visible HAeBe stars are marked as stars,
while the embedded millimeter source is indicated by the square.
Contours illustrate the emission to the blue ($-4$ to $-1$ \kms) and
to the red (1 to 4 \kms) of the peak.  The contour levels range from
5.2 to 26 \jybeam\ \kms\ in steps of 2.6 \jybeam\ \kms.  The
resolution is 6.75\arcsec\ $\times$ 5.5\arcsec. At 600 pc, the beam
corresponds to a linear scale of 4000 $\times$ 3300 AU.  The box
indicates the field of view of the uniform-weighted image at
right. {\it Right:} A high resolution datacube is produced from
uniformly-weighted BIMA array data combined with FCRAO data. The
greyscale shows zeroth moment over the entire range of CO emission
($-4.0$ to 4.2 \kms). The blue and red contours show the zeroth
moments over the ranges from $-4.0$ to $-2.1$ \kms\ and 2.3 to 4.2
\kms, respectively.  The contour levels range from 1.8 to 4.2 \jybeam\
\kms\ in steps of 0.6 \jybeam\ \kms.  The resolution of 3.4\arcsec\
$\times$ 2.4\arcsec\ (2040 $\times$ 1440 AU) highlights primarily
small features in the outflowing gas. This map shows that no
significant red- or blue-shifted emission is associated with the
\lkha\ 198-IR source.  For comparison, the polarized intensity in
H-band as measured by \cite{per04} is shown in the inset. The \lkha\
198 and \lkha\ 198-IR sources are marked on this image.}
\label{198co_mom0}
\end{figure*}

\section{\lkha\ 198 and Environs}
\label{lkha198}

\lkha\ 198 (V* V633 Cas, IRAS 00087+5833) is an HAeBe star located at
$\sim 600$ pc based on the interstellar extinction law
\citep{rac68,cha85}.  \cite{hil92} categorize \lkha\ 198 as an A5
pre-main sequence star while \cite{her04} find a spectral type of B9
$\pm 2.5$.  \lkha\ 198 is classified as a Group II HAeBe star in the
nomenclature of \cite{hil92}, meaning that the SED rises with
wavelength in the infrared. Group II HAeBe stars have significant
circumstellar envelopes which reprocess stellar radiation at long
wavelengths. A strong CO bipolar outflow is associated with this
region, but it is not centered on \lkha\ 198, but rather to the
northwest \citep{can84}. \cite{str86} detected evidence of a
blue-shifted optical flow extending to the east-southeast of \lkha\
198, which did not coincide with the red-shifted CO emission in that
direction.  An infrared companion (\lkha\ 198-IR) was identified
approximately 6\arcsec\ north of \lkha\ 198 by \cite{lag93}. A
submillimeter source (\lkha\ 198-mm) was identified 19\arcsec\
northwest of \lkha\ 198 at 800 \micron\ by \cite{san94} who also found
evidence for CO outflows in the three sources \lkha\ 198, \lkha\ 198
mm and V376 Cas. This source is another HAeBe star in the region,
located about 40\arcsec\ north of \lkha\ 198.  \lkha\ 198 mm was also
strongly detected in a 1.3 mm observation of the region by
\cite{hen98}.  High resolution interferometry toward \lkha\ 198 and
the surrounding region at 2.7 mm failed to detect \lkha\ 198, but
\lkha\ 198-mm was detected and resolved \citep{dif97}.  High
resolution laser guide star adaptive optics imaging polarimetry by
\cite{per04} revealed a centrosymmetric polarimetric pattern
consistent with a conical cavity. They also detected a ``cometary''
structure in JHK intensity images.  Their data suggest that this
structure is associated with \lkha\ 198-IR.

\begin{figure*}
\vspace*{10cm}
\includegraphics{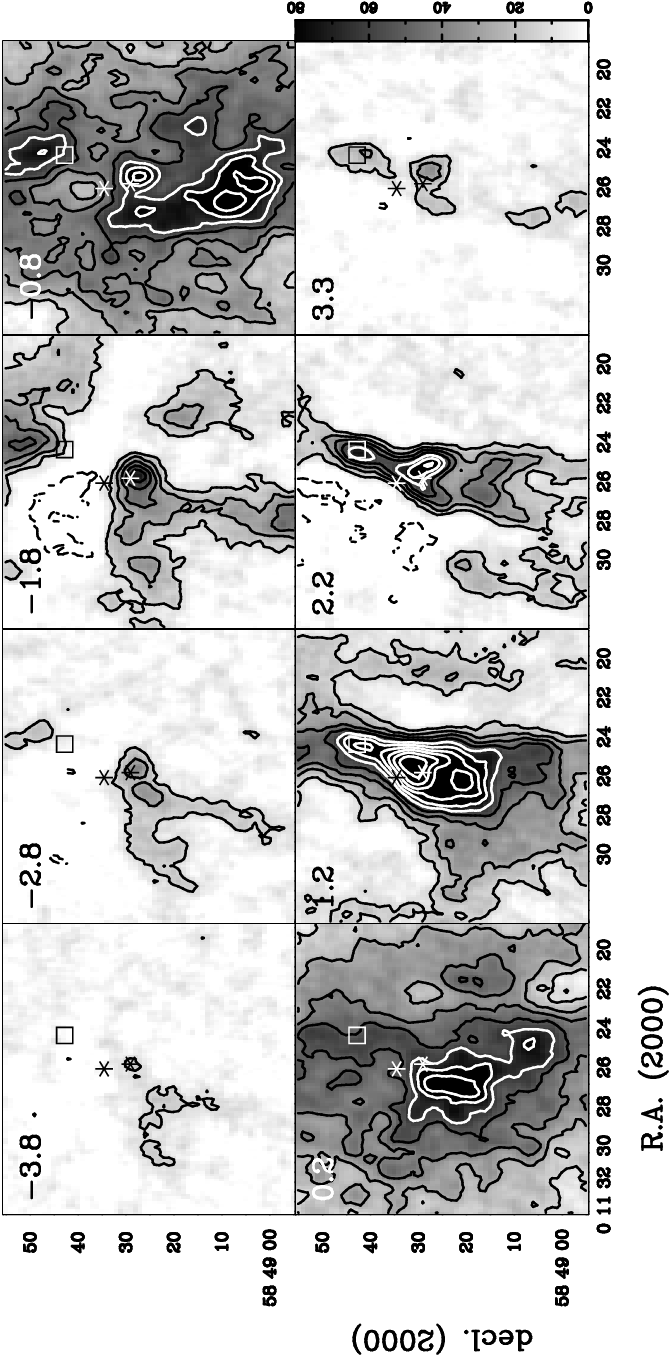}
\caption{Binned intensity images of \co\ from combined FCRAO and BIMA
array data toward the \lkha\ 198 region is shown in greyscale and
contours over 1.016 \kms\ bins. The mean velocity per channel is
noted.  The greyscale range is common to all channels and is shown at
right (\jybeam). The contours are plotted at 20 to 200 $\sigma$ in steps of 20
$\sigma$ where $\sigma = 0.64$ \jybeam\ is the rms level in a single
channel before binning.  Sources are marked by asterisks and the
square as in Figure \ref{198co_mom0}.}
\label{198co_channel}
\end{figure*}

\subsection{Observational Results}

\subsubsection{Moment and Channel Maps} 

Figure \ref{198co_mom0} shows the zeroth moment of the combined map of
\co\ emission toward the \lkha\ 198 region from the combination of
channels from $-4$ to 4 \kms\
and the distribution of emission at velocities blueward and redward of
the systemic LSR source velocity, $-0.7$ \kms\ \citep{can84}.  The
moment maps were created by using only channels with flux densities
exceeding 1 \jybeam, approximately 1.5 times the rms level per
channel.  The velocity ranges over which they are taken are specified
in Figure \ref{198co_mom0}. The resolution in this map is an order of
magnitude improved over the \co\ maps of \cite{can84} but covers an
area limited by the primary beam of the BIMA array antennas.  Strong
\co\ emission is concentrated around \lkha\ 198 rather than \lkha\
198-IR  to the north.  We have not detected any structures
similar to the dust extension from \lkha\ 198-IR observed in near-IR
polarimetry data \citep{per04}.

The second panel of Figure \ref{198co_mom0} shows a higher resolution
image of the region surrounding \lkha\ 198 and \lkha\ 198-IR produced
by uniformly weighting the BIMA array data before combination with the
FCRAO map.  The CO features in this map are concentrated around \lkha\
198, with no significant emission detected around \lkha\ 198-IR.
Blue- and red-shifted emission peaks are paired on either side of
\lkha\ 198.  If these each represent individual outflows driven by
independent sources, then the driving sources are not evident as no
sources of strong dust emission are located at the central positions
between the red and blue peaks. Instead, it is more likely that the
peaks represent parts of an outflow from \lkha\ 198 which may be
precessing over time.  Such precession can occur when the outflow's
driving source is a member of a multiple system.  \cite{smi05} show
that \lkha\ 198 is a spectroscopic binary with a separation of 60 mas.

Figure \ref{198co_channel} shows binned channel maps with 1 \kms\
steps around the V(LSR) velocity of $-0.7$ \kms.  In the blue-shifted
channels, the dominant feature is the arc of emission that traces out
the edges of the southeast elliptical cavity detected in the I-band
image of \cite{lag93} and the J-band image of \cite{kor97}.
This feature appears more strongly focussed on \lkha\ 198 than the
adjacent \lkha\ 198-IR.  A secondary blue-shifted feature to
the north (extending off the map area) may be a component of the
outflow originating at \lkha\ 198-mm. The channels around the cloud's
velocity are strongly dominated by cloud emission, but the peak
associated with \lkha\ 198 persists over most channels.  By the 1.2
\kms\ channel, a red-shifted outflow component can be associated with 
\lkha\ 198-mm. The red-shifted counterpart to the \lkha\ 198 outflow
appears along the same axis. At the highest velocity shown, 3.3 \kms,
components of the two outflows can be distinguished.  Interestingly,
red-shifted emission to the south and southeast also seems to trace
out the elliptical cavity detected by \cite{lag93} in the optical.
Interestingly, we find no strong \co\ features associated with \lkha\
198-IR.

\begin{figure*}
\vspace*{10cm}
\includegraphics{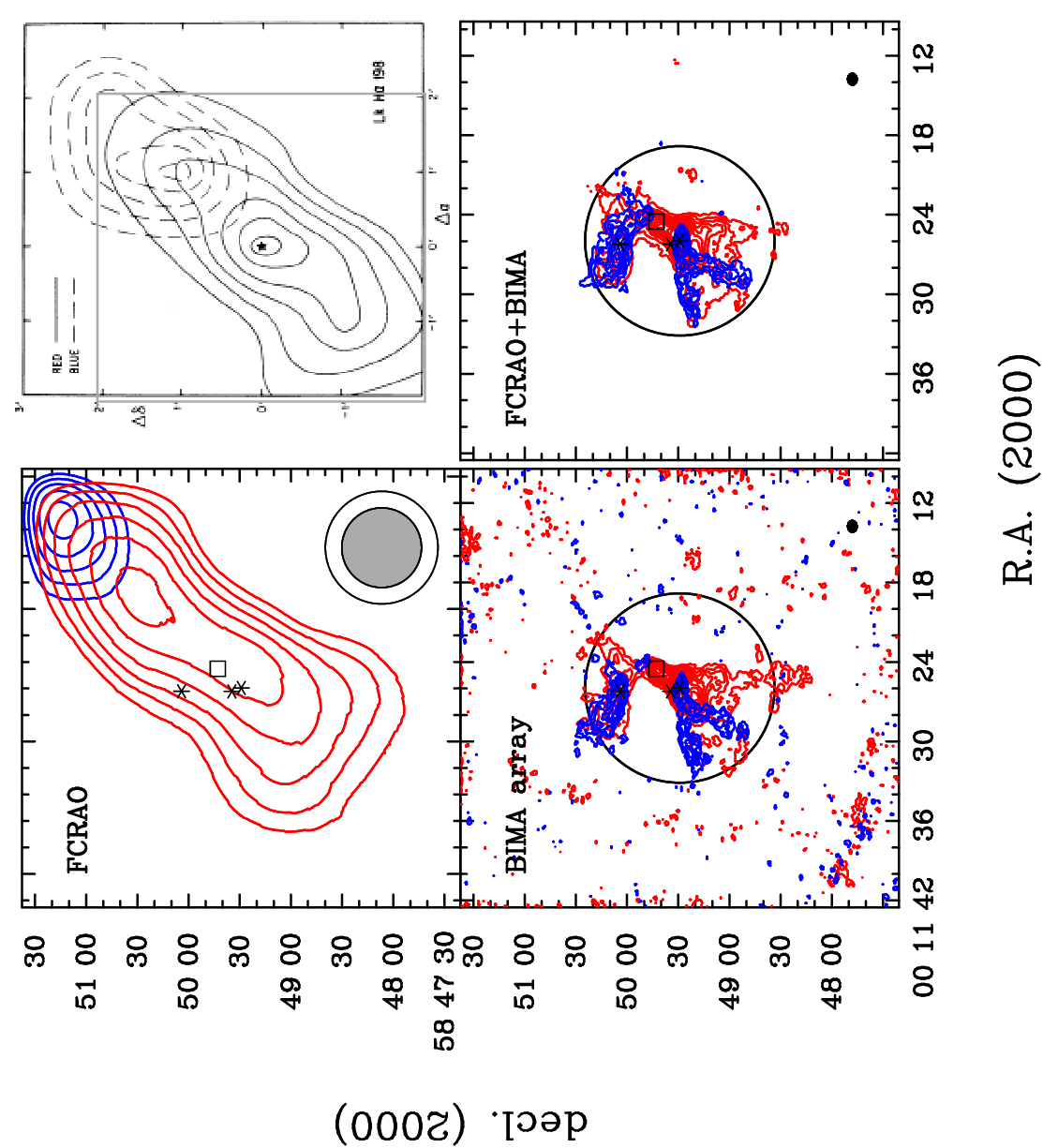}
\caption{Zeroth moment maps of the \lkha\ 198 region in \co\ from
FCRAO (top left), the BIMA array (bottom left) and the combined
dataset (bottom right). Figure 3 of \cite{can84} is reproduced at the
top right. The grey box shows the area covered by our maps.  The
contours of the FCRAO image are produced by integration over the same
velocity channels as Canto et al.\ (i.e., $-4.7$ to $-1.5$ \kms\ for the
blueshifted emission and 0.7 to 3.9 \kms\ for the redshifted emission)
and plotted at comparable levels (312 \jybeam\ to 468 \jybeam\ in
steps of 52 \jybeam).  Channels with flux densities less than 35
\jybeam\ were not included. The elongated red lobe dominates the
emission. The blue lobe of \cite{can84} is just detected at the NW
edge of our map, which is less extensive in that direction.  The
resolution of the FCRAO data is 46.7\arcsec\ (grey circle), which we
have convolved to 66\arcsec\ (white circle) to equal the resolution of
the \cite{can84} map.  The BIMA array data are integrated over the
same velocity ranges, with only channels with flux densities exceeding
1 \jybeam\ included.  The circle denotes the primary beam of BIMA at
115 GHz.  This is effectively the edge of sensitivity in the BIMA and
combined maps.  The combined FCRAO+BIMA map is made with the same
contour levels as shown in Figure \ref{198co_mom0} and integrated over
the same velocities.  The BIMA array and combined maps have a
synthesized beam of 6.75\arcsec\ $\times$ 5.5\arcsec\ as indicated at
the bottom right of those maps.  The sources are indicated by
asterisks and the square as for Figure \ref{198co_mom0}.}
\label{198co_FCRAO_BIMA}
\end{figure*}

We note that Figure \ref{198co_mom0} may seem initially at odds with
the results of Figure 3 of \cite{can84} who report single dish
detection of \co\ in a 66\arcsec\ beam.  The red lobe peak was found
to extend to the southeast and to be centered on \lkha\ 198 with a
1.5\arcmin\ offset to the blue peak to the northwest. \cite{can84} did
note that their observations did not depict lobes symmetric about the
star, as is more typical for a bipolar outflow, but they did interpret
the blue and red shifted emission as an outflow driven by \lkha\ 198.
It is noteworthy as well that blue-shifted optical emission was
detected to the southeast of \lkha\ 198 by \cite{str86}.  Our
single-dish FCRAO data have higher angular resolution (46.7\arcsec)
than the \cite{can84} map.  We have therefore convolved our FCRAO data
to 66\arcsec\ resolution (see Figure \ref{198co_FCRAO_BIMA}).  When
compared to the existing dataset, we find a similar morphology in the
zeroth moment map, with a single prominent red lobe to the southeast
and a blue lobe to the northwest.  The FCRAO plot in Figure
\ref{198co_FCRAO_BIMA} has been generated using the same range in
velocity as Figure 3 of Canto et al.\ and plotted with comparable flux
density levels.  At a resolution of 66\arcsec, the emission in the
region is strongly dominated by red-shifted emission which is centered
on the \lkha\ 198/198-IR sources. The blue-shifted emission in the map
lies approximately 1\arcmin\ to the north \citep[as found by
][]{can84}. The fact that the outflow has been acknowledged to be very
asymmetric about the attributed driving source, either \lkha\ 198 or
\lkha\ 198-IR \citep{can84,nak90}, could point to multiple driving
sources, but could also be explained if the outflow is not much
inclined from the line of sight.

The ambiguity regarding the asymmetric nature of the outflow is resolved
by the zeroth moments of the BIMA array data alone, which are also shown in
Figure \ref{198co_FCRAO_BIMA}. This map illustrates sharply the
effect of a ten-fold improvement in resolution on the interpretation
of the number and orientation of outflows in a region containing
multiple sources. Even though we are only sensitive to structures on
limited spatial scales with these data, the features in red and blue
emission can be more clearly associated with individual objects than
is possible based on single dish observations. Figure \ref{198co_FCRAO_BIMA}
shows that the red lobe resolves into distinct features, likely
associated with \lkha\ 198 and/or \lkha\ 198-IR, a distinct red peak
associated with \lkha\ 198-mm, and larger scale blue- and red-shifted
emission in the vicinity of V376 Cas.  Whether the blue-shifted
emission to the southeast is driven by \lkha\ 198 or \lkha\ 198-IR
cannot be easily distinguished in this map, since their small
separation of 5\arcsec\ is comparable to our resolution. However, the
channel maps of Figure \ref{198co_channel} do favour \lkha\ 198 as the
driving source for both outflows originating in that small area.  The
improved resolution of our data explains the coincidence of the
blue-shifted optical flow defined by the [SII] emission of
\cite{str86} within the red-shifted CO outflow lobe detected by
\cite{can84}.

With the higher resolution data in hand, we can interpret the single
dish data as follows.  The red-shifted emission originating
from the \lkha\ 198/198-IR objects completely dominates the emission
near the four sources, with the weaker blue lobe shifted north
because it originates from one or both of V376 Cas or \lkha\ 198-mm.
This general \co\ morphology in Figure \ref{198co_mom0} is discernable
from the CO $J=2-1$ emission of \cite{san94} observed with
20\arcsec\ resolution with the JCMT, which shows that blue-shifted and
red-shifted emission overlap southeast of the \lkha\ 198/198-IR
sources and that secondary blue and red-shifted peaks are seen
$> 1$\arcmin\ north of these objects.  However, with the
superior resolution of the BIMA array (and the removal of large scale
emission from the maps), the outflow structure can be resolved.

\begin{figure}
\vspace*{6.5cm}
\includegraphics{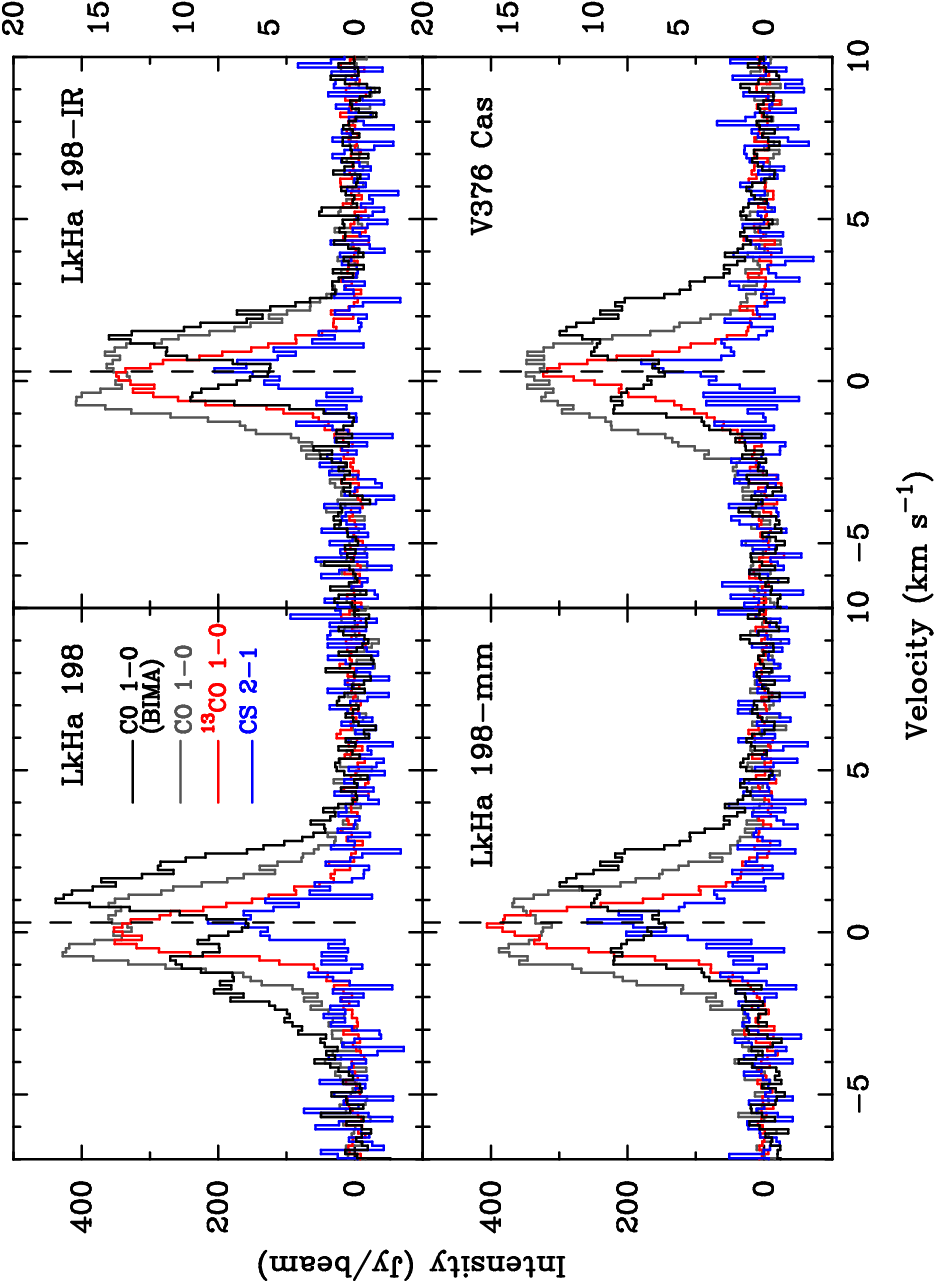}
\caption{Spectra toward individual sources in the \lkha\ 198 region
  from FCRAO and FCRAO+BIMA array data. The spectral resolution is
  0.127 \kms.  Grey, red and blue spectra represent \co, \tco\ and
  \cs\ observations from FCRAO which were regridded to the spatial and
  spectral scales of the FCRAO+BIMA \co\ map.  The intensity scale at
  left is for the single dish \co\ emission.  The \tco\ and \cs\
  emission have been multiplied by factors of two and twenty.  The
  black spectra are taken from the combined \co\ map. The intensity
  scale is shown at right.  Self-absorption is noted toward each
  source in the field in both the single dish and combined \co\
  spectra.  The strongest evidence of high velocity line wings is
  observed toward \lkha\ 198 and \lkha\ 198-mm.  Some red-shifted
  emission is observed toward \lkha\ 198-IR. }
\label{198_spectra}
\end{figure}

\subsubsection{Spectra}

In addition to the images presented above, we can construct spectra
toward the known point sources in the field, including \lkha\ 198.
Such spectra can indicate whether line emission is optically thick or
self-absorbed (from the line shape) or if high-velocity gas is
associated with a given position (though the presence of emission in
the line wings).

Figure \ref{198_spectra} presents the \co\ high spatial and
spectral resolution data from the combined FCRAO and BIMA array data
cubes toward each of the four sources in the \lkha\ 198 region.  The
spectral resolution of 0.127 \kms\ is a modest factor of two improvement from
the work of \cite{can84}. Strong self-absorption is evident toward
each of the HAeBe objects of Figure \ref{198_spectra},
indicating that each remains embedded within the nascent molecular
cloud. Line wings are detected toward both \lkha\ 198 and \lkha\
198-mm, while there is a suggestion of some red-shifted higher
velocity emission toward \lkha\ 198-IR, although this feature could be
associated with the ambient cloud material since it arises in the
combined spectra, but is not apparent in BIMA array data alone.  

Our FCRAO spectral line data enables us to assess whether tracers of
dense ($N(H_2) \sim 10^5$ cm$^{-3}$) gas such as \cs\ and \nthp\ are
present in the environment around \lkha\ 198.  Figure
\ref{198_spectra} shows \co, \tco\ and \cs\ emission from FCRAO
(integrated from a regridded map to match the BIMA array pixel scale,
map size and velocity resolution). \nthp\ was not detected anywhere in
the observed field. The \cs\ emission is peaked toward the
self-absorption of the spectra derived from the high spatial
resolution combined data. This is not surprising, since the \cs,
despite the limited spatial resolution of the maps, indicates the
presence of relatively high density material ($\sim 10^4$ cm$^{-3}$).
For the two northern sources, the \cs\ peak also aligns well with
the peak of the \tco\ and the absorption feature of single dish \co,
at 0.3 \kms.  For the two southern sources, the \tco\ peak appears
somewhat blue-shifted compared to the \cs\ peak, with the greatest
offset toward \lkha\ 198.

Ideally, we could use the high spatial resolution to solve for the
V(LSR) as a function of position across the region. However, the \co\
emission spectra are too complex to be easily fit for single V(LSR)
solutions.  The \cs\ emission would be ideal, but for this species we
have only FCRAO data.  These data show no significant variation in
peak velocity position across the FCRAO maps (i.e., 4\arcmin).

A comparison of the \co\ spectra from the single dish alone and the
combined map (notably with different beams) indicates that the line
wings are much more prominent in the spectra in which the BIMA array
data are included.  High resolution data enable us to discriminate
between sources, e.g., clarifying the absence of line wings toward \lkha\
198-IR.  In addition, these data have increased sensitivity to outflow
features which are significantly beam-diluted in the single dish maps.
\lkha\ 198 has extensive line wings shifted blueward and redward of
the source's velocity. Toward \lkha\ 198-mm and V376 Cas, the
red-shifted emission is more prominent than the blue-shifted emission
in the combined data but the opposite appears to be true in the single
dish data, perhaps indicating that the blue lobe of the outflow is
more diffuse than its red counterpart. We also note  that evidence for
a blue wing is apparent in the \tco\ spectrum toward V376 Cas. 

\subsubsection{2.6 mm Continuum}

No 2.6 mm continuum emission was detected toward the \lkha\ 198
region.  This is consistent with the absence of a detection at 2.7 mm
of dust emission toward \lkha\ 198 and V376 Cas by \cite{dif97}.
Even in regions of \co\ and \nthp\ emission, no continuum emission is
evident down to a threshold of 57 \mjybeam, which is our 3 $\sigma$
limit.  \cite{dif97} did detect 2.7 mm emission toward \lkha\ 198-mm
using the Plateau de Bure IRAM interferometer. They detected a peak
flux density of $S_\nu = 13$ \mjybeam\ with a resolution of
9.7\arcsec\ $\times$ 5.1\arcsec\ and an integrated flux density of $24
\pm 6.9$ mJy.  Our rms level of 18.5 \mjybeam\ is insufficient to
detect this flux density.  Our lack of detections also indicates that
there is no significant free-free emission in the region.

\subsection{Individual Objects}

We summarize some general findings toward each source in our two
fields in Table \ref{sources_lkha198}.  Some sources are
associated with CO emission, but not definitively with an outflow. All
four sources in the \lkha\ 198 field are associated with local \co\
emission, as revealed by the strong self absorption in each of the
spectra of Figure \ref{198_spectra}.  \tco\ and \cs\ are also detected
toward each of the sources. 

\begin{deluxetable*}{llllcccc}
\tablecolumns{8} 
\tablewidth{0pc} 
\tablecaption{\lkha\ 198 \& Surrounding Objects}
\tablehead{\colhead{Source} & \colhead{Other ID} & 
\colhead{RA} & \colhead{DEC} & \colhead{Group$^a$} &
\colhead{Associated} & \colhead{\# Associated} & \colhead{PA} \\
& & \colhead{(J2000)} & \colhead{(J2000)} & & \colhead{CO} &
\colhead{Outflows} & \colhead{($^\circ$ E of N)} }
\startdata 
\lkha\ 198 & V633 Cas & 00:11:25.97 & +58:49:29.1 & II & yes & 2 &
120\degr, 70\degr\ \\
\lkha\ 198-IR & \lkha\ 198 B & 00:11:26.22 & +58:49:34.58  & ... & yes & none & -- \\
\lkha\ 198-mm &  & 00:11:24.51 & +58:49:42.81 & ... & yes & 1 &
350\degr\ \\
V376 Cas & & 00:11:26.21 & +58:50:04.1 & II & yes & 1 & 70\degr\ \\
\enddata 
\tablecomments{$^a$ Classification of \cite{hil92}.}
\label{sources_lkha198}
\end{deluxetable*}

\subsubsection{\lkha\ 198}

Strong evidence for a blue-shifted cavity centered on \lkha\ 198 at a
position angle\footnote{We define the position angle according to the
component of the outflow which is most clearly defined.}  of $\sim
120^\circ$ is evident in the integrated channel maps of Figure
\ref{198co_channel}. The same emission cavity is also obvious in the
BIMA array data of Figure \ref{198co_FCRAO_BIMA}.  This emission is
likely associated with the cavity traced by near-IR I-band emission
\citep{cor95,lag93,lei91}.  The zeroth moment map of Figure
\ref{198co_mom0} also highlights a weaker blue-shifted feature to the
southwest of the \lkha\ 198 sources (at a position angle of $\sim
250^\circ$ E of N). This could be a portion of a second outflow
centered on \lkha\ 198.

The appearance of a single extensive blue lobe to the northwest and
driven by \lkha\ 198 is shown to be a result of resolution
limitations in earlier data.  \cite{nak90} indicated that the velocity
width of the blue lobe was significantly larger than the width of the
red lobe they detected. Their single blue lobe is a combination of the
blue-shifted emission driven from V376 Cas and the young, embedded
source \lkha\ 198-mm. However, the double-peaked, red-shifted emission
of \cite{nak90} persists even in our higher spatial resolution
data. The peak observed to the north of \lkha\ 198 may arise from red-shifted
emission from a \lkha\ 198-driven outflow in combination with emission
from an outflow driven by \lkha\ 198-mm, while the red-shifted emission
to the south of \lkha\ 198 indeed arises from an outflow driven by
that source.  We suggest that at least one outflow is being driven
from \lkha\ 198, based on the extended blue-shifted cavity.  
\lkha\ 198-IR has been identified as the source associated with the
cavity by \cite{cor95}, but our \co\ observations suggest that \lkha\
198 is driving at least one outflow to the southeast.

Figure \ref{198co_mom0} compares the high resolution \co\ data with
the polarized intensity at H-band measured by \cite{per04}.  The
bipolar nebulosity centered on \lkha\ 198 in H-band emission has no
counterpart in \co\ from BIMA array data. In fact, the high resolution
components of the red- and blue-shifted \co\ emission is oriented
primarily east and west of \lkha\ 198.  The polarized emission should locate the
circumstellar dust around the star; in this case, the north-south
alignment could indicate the region of the star where the envelope has
not yet been eliminated by the action of the outflow(s).  

\subsubsection{\lkha\ 198-IR}

Based on Figure \ref{198co_mom0}, there is little evidence for any
significant outflow centered on \lkha\ 198-IR.  Figure
\ref{198_spectra} shows that there is no high velocity blue wing
associated with the infrared source and only a suggestion of slight
red-shifted emission. However, like \lkha\
198, \lkha\ 198-IR shows strong self-absorption of the CO line,
meaning that circumstellar gas is still associated with this object. 

The ``cometary'' structure detected in intensity and polarization in
the near-IR has no detected counterpart in \co\ emission.  
It is difficult to reconcile the HH
objects A and B from \cite{cor95} as originating from \lkha\ 198-IR
without any hint of a \co\ outflow in our data.  In fact, there is a
hint of a red-shifted line wing in the \co\ spectrum toward \lkha\
198-IR (see Figure \ref{198_spectra}). This could indicate a
large-scale outflow contribution or ambient material contributions to
the spectrum. 

\subsubsection{\lkha\ 198-mm}

A roughly north-south outflow (with blue-shifted emission at a
position angle of $\sim 350^\circ$) from \lkha\ 198-mm
is strongly suggested by Figures \ref{198co_mom0} and
\ref{198co_channel}.  The blue-shifted component is particularly
apparent in Figure \ref{198co_channel} and appears quite localized to
the source. This is also suggested by the BIMA array data alone in
Figure \ref{198co_FCRAO_BIMA} although we note that there is
significant emission missing from this map.  The red lobes of the
outflows of \lkha\ 198 and \lkha\ 198-mm clearly overlap.  It is
certain that part of the \co\ red-shifted emission detected by
\cite{can84} originates from this source and not \lkha\ 198.

Figure \ref{198_spectra} shows the self-absorption present in all
sources in the region, indicating that substantial circumstellar gas
remains. Figure \ref{198_spectra} also shows that the red lobe from
\lkha\ 198-mm is much stronger than the blue lobe. The fact that the
outflow emission appears to lie so close to the driving source may
indicate either that the outflow is almost entirely along the line of
sight or that the driving source is very young.  The absence of very
high velocity line wings argues against a line-of-sight
alignment. Superior resolution will be required to disentangle the
outflow of \lkha\ 198-mm from the other outflows in the region and the
ambient cloud. We note that the orientation of the outflow is
completely consistent with a series of HH objects observed at position
angles of $160^\circ$ and $340^\circ$ \citep{cor95} which were
attributed to \lkha\ 198, but are in fact more symmetric about \lkha\
198-mm.

\begin{figure*}
\vspace*{9.5cm}
\includegraphics{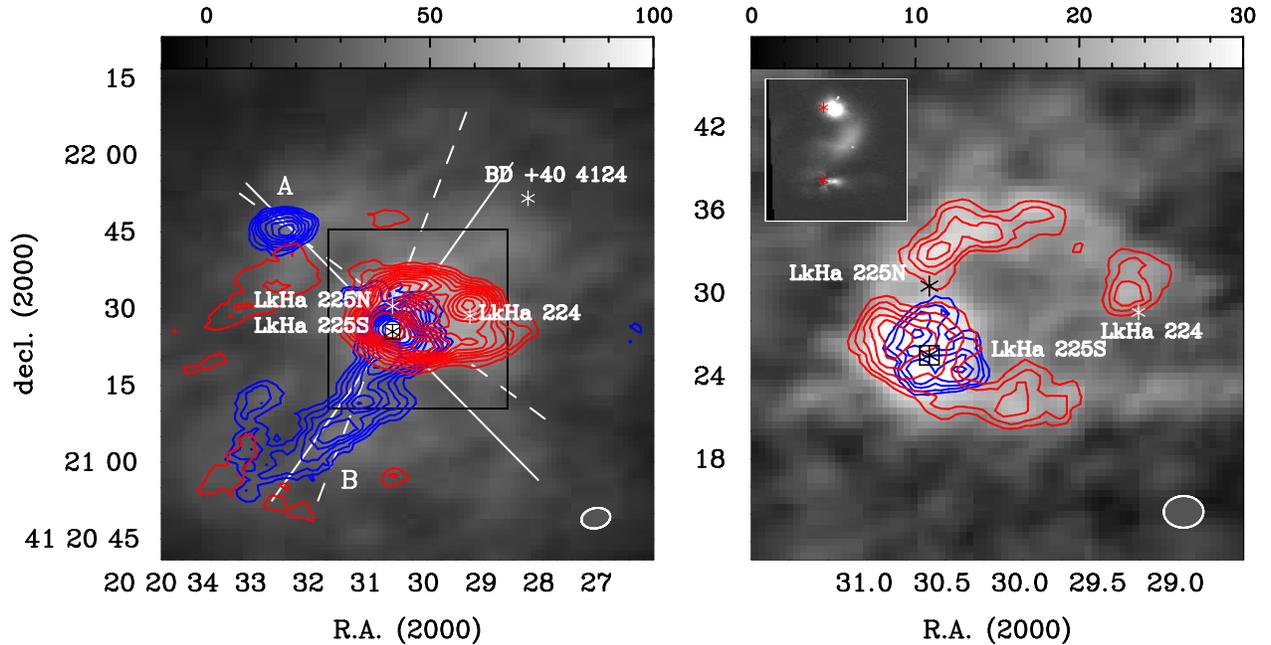}
\caption{{\it Left:} The greyscale is the zeroth moment over the
entire range of CO emission (0 to 18 \kms) of the combined FCRAO and
BIMA array data toward the region centered on \lkha\ 225S (in units of
\jybeam\ \kms).  Optically visible HAeBe stars are marked by stars,
while the embedded submillimeter source is indicated by a square.
Contours trace blue emission from 0 to 4 \kms\ and red emission from
12 to 18 \kms.  The blue contour levels range from 4.5 to 14.4 \jybeam
\kms\ in steps of 0.9 \jybeam\ \kms. The red contour levels range from
6.3 to 24.3 in steps of 1.8 \jybeam\ \kms. The resolution is
5.7\arcsec\ $\times$ 4.0\arcsec, or 5700 $\times$ 4000 AU at 1 kpc.
Two distinct blue-shifted features, ``A'' and ``B'' are observed.
Position-velocity slices are taken through both features to align with
the \lkha\ 225S (solid lines) and \lkha\ 225N (dashed lines)
sources. The box shows the field of view of the uniform-weighted image
at right. {\it Right:} A combined map generated with the
uniformly-weighted BIMA array data yields a resolution of
2.87\arcsec\ $\times$ 2.29\arcsec. The symbols and velocity ranges of
the zeroth moment maps are the same as for the naturally-weighted
image. The blue contour levels range from 5 to 10 \jybeam\ \kms\ in
steps of 1 \jybeam\ \kms, and the red contour levels range from 10 to
20 \jybeam\ \kms\ in steps of 2 \jybeam\ \kms. For comparison, the polarized intensity in
H-band as measured by \cite{per04b} is shown in the inset. The \lkha\
225S and \lkha\ 225N sources are marked on this image.}
\label{225co_mom0}
\end{figure*}

\subsubsection{V376 Cas}

\cite{hil92} categorize this source as a Group II HAeBe star, meaning
that its SED is flat or rising toward longer wavelengths. We therefore
expect that there is a substantial amount of cool material around the
star which reprocesses more of the total radiation than could be
explained by a flat disk. It is therefore not surprising that the \co\
spectrum of this source shows evidence of self-absorption in Figure
\ref{198_spectra}.  Evidence for nebulosity has been previously noted
to the west of V376 Cas by several authors
\citep{lei91,pii92,smi04}. \cite{lei91} used polarized imaging techniques to
detect the elongation of this nebulosity ($120^\circ$) and clearly
identify V376 Cas as the illumination source of the nebula.

Figure \ref{198co_mom0} demonstrates the presence of extended,
overlapping blue- and red-shifted emission around this source at a
position angle of $\sim 70^\circ$. The BIMA array data alone, shown in
Figure \ref{198co_FCRAO_BIMA}, suggest that the strong blue-shifted
emission peak noted northwest of \lkha\ 198 in single-dish maps of the
region may be dominated by emission from V376 Cas.  This outflow
orientation seems consistent with the nebulosity observed in the
optical, given the difference scales to which those data and our
molecular lines maps sample. The blue outflow is also coincident with
the known HH objects to its east, designated HH162 \citep[see Fig.\ 1
of][]{cor95}. The dominance of the blue-shifted emission to the east
and the lack of comparable red-shifted emission to the west is
entirely consistent with V376 Cas being located at the eastern edge of a
small dark cloud core \cite[see Fig.\ 1 in][]{smi04}.

\section{\lkha\ 225S and Environs}
\label{lkha225S}

Another example of an HAeBe star in a group is \lkha\ 225.  \lkha\ 225
comprises two sources separated by approximately 5\arcsec\
\citep{asp94}: \lkha\ 225N and \lkha\ 225S (V* V1318 Cyg N and V*
V1318 Cyg S) at a distance of 1 kpc (\citealt{can84} after
\citealt{her60}).  At wavelengths longward of 5 \micron, \lkha\ 225 is
the dominant source in the region \cite[$L \sim 1600 L_\odot$,
][]{the94}.  A third peak observed by \cite{asp94} in the optical
arises due to nebulosity between the two sources.  This peak is
coincident with the comma shaped reflection nebula extending
from \lkha\ 225N to \lkha\ 225S detected by \cite{per04b}. \cite{pal95}
detect two sources in the K-band with no evidence of the third
source noted by \cite{asp94} and also detect an H$_2$O maser
coincident with \lkha\ 225S lying at the center of a CO outflow.  The
detection of continuum peaks at the position of \lkha\ 225S at 1.3 mm
\citep{hen98} and in high resolution imaging at 2.7 mm and 3.1 mm
\citep{dif97,loo06} confirm that \lkha\ 225S is deeply embedded and
likely the youngest star in the region. \lkha\ 225N is likely a
pre-main sequence star.  Near-IR stellar spectra \citep{dav01} show no
absorption lines, only emission lines, making the identification of
the spectral types of both stars difficult.  Two other well-studied HAeBe stars are
in the same region: BD +40 4124 (V1685 Cyg, MWC 340, IRAS 20187+4111)
and \lkha\ 224 (V1686 Cyg), located approximately 50\arcsec\ to the
northwest and 15\arcsec\ to the west, respectively, from the \lkha\
225 sources.  \cite{loo06} show that the \lkha\ 225 sources lie in the
brightest molecular cloud core (as traced by BIMA array observations
of \cs) in an elongated molecular filament.  \cs\ emission at
2.5\arcsec\ resolution is strongly peaked on \lkha\ 225S.

\begin{figure*}
\vspace*{10cm}
\includegraphics{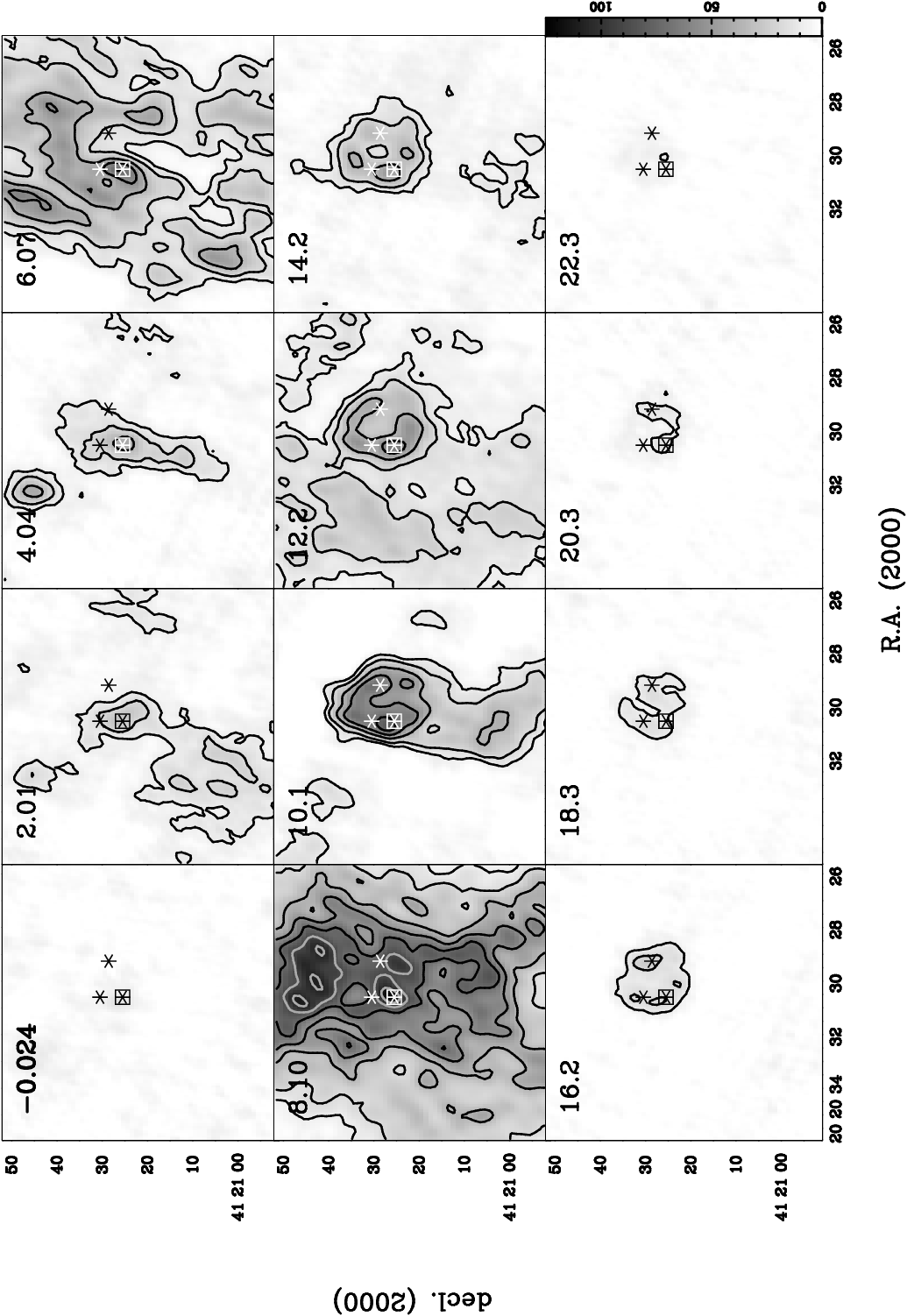}
\caption{Binned velocity channels of \co\ toward the \lkha\ 225S
  region are shown in greyscale and contours over 2.03 \kms\ bins. The
  mean velocity per channel is noted in \kms.  The greyscale range is
  common to all channels and is shown at right (\jybeam). The contours
  are plotted at 20 and 40 to 360 $\sigma$ in steps of 40 $\sigma$
  where $\sigma = 0.45$ \jybeam\ is the rms level in a single channel
  before binning.  The sources are marked by asterisks and a square as
  in Figure \ref{225co_mom0}.}
\label{225co_channel}
\end{figure*}

\subsection{Observational Results}

\subsubsection{Moment and Channel Maps}

Figure \ref{225co_mom0} shows the zeroth moment map summed from 0 to
15 \kms, and the distribution of emission blueward and redward of the
source's LSR systemic velocity, 7.9 \kms\ \citep{pal95}.  The \co\
emission peaks around \lkha\ 225S. This concentration is unsurprising
since a submillimeter peak is also detected at its position,
indicating that the associated source is still deeply embedded and
very young.  The uniform-weighted image shows clearly that the CO
peaks do not coincide with the positions of the potential driving
sources, indicating that the \co\ emission associated with these
sources is primarily in outflowing gas.  Figure \ref{225co_channel}
shows channel maps of \co\ emission. The \co\ morphology is very
complex as evidenced by the variation of structure with velocity. The
most intriguing structure is an apparent ring of emission, centred at
$\alpha_{J2000} = 20^{\rm h}20^{\rm m}$30\fs0, $\delta_{J2000} =
+41$\degr21\arcmin28\farcs0, which persists over a large range of
velocity (10 to 18 \kms). A SIMBAD\footnote{Guest User, Canadian
Astronomy Data Centre, which is operated by the Dominion Astrophysical
Observatory for the National Research Council of Canada's Herzberg
Institute of Astrophysics.}  search lists no objects at the center of
the ring; the closest objects are the HAeBe stars in the field. An
H$_2$O maser is reported at the position of \lkha\ 225S, identified as
the driving source of the $^{12}$CO $J=2-1$ outflow seen by \cite{pal95}
at 15\arcsec\ resolution.  The blue-shifted emission detected by
\cite{pal95} lies just to the east of the peak of the red-shifted
emission, implying that the outflow lies almost along the
line-of-sight.  Peaks on the ring are coincident with the positions of
\lkha\ 224 and \lkha\ 225S, but \lkha\ 225N lies between \co\
peaks. We surmise that the ring is the supposition of emission from
several outflows, dominated by the pole-on outflow, which could have a
substantial opening angle and could appear as a cylindrical cavity.
No obvious counterpart to the ring is observed in blue-shifted
emission, but strong blue-shifted emission is concentrated at the
position of \lkha\ 225S.

\begin{figure}
\vspace*{7.3cm}
\includegraphics{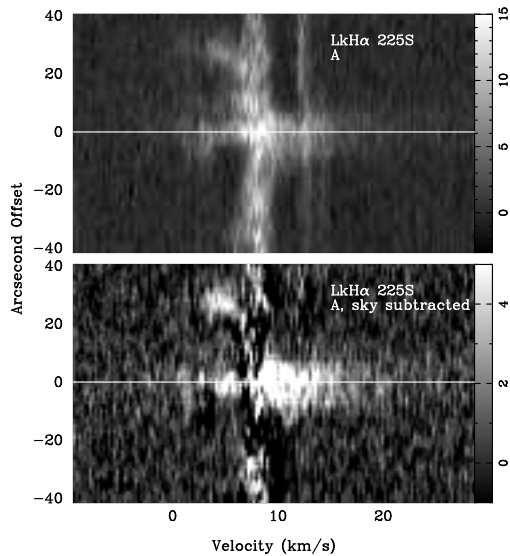}
\caption{The top panel shows a position velocity diagram taken centred
on \lkha\ 225S sliced through the blue-shifted feature ``A'' of Fig.\
\ref{225co_mom0}.  The evidence for multiple clouds along the line of
sight is indicated by the bands of emission at all angular positions.
The bottom panel shows the same position velocity slice, but with a
second slice, positioned off the source 24\arcsec\ to the southeast,
subtracted from the on-source slice. The bulk of the cloud emission is
removed, especially the fainter clouds. In this image, the most
prominent feature is the emission extending from about 0 \kms\ to 20
\kms\ at zero angular offset. We interpret this as evidence for a
pole-on outflow, most likely centered on \lkha\ 225S.  Blue-shifted
emission is also detected to the northeast of \lkha\ 225S.}
\label{pv_225s}
\end{figure}

\begin{figure*}
\vspace*{10cm}
\includegraphics{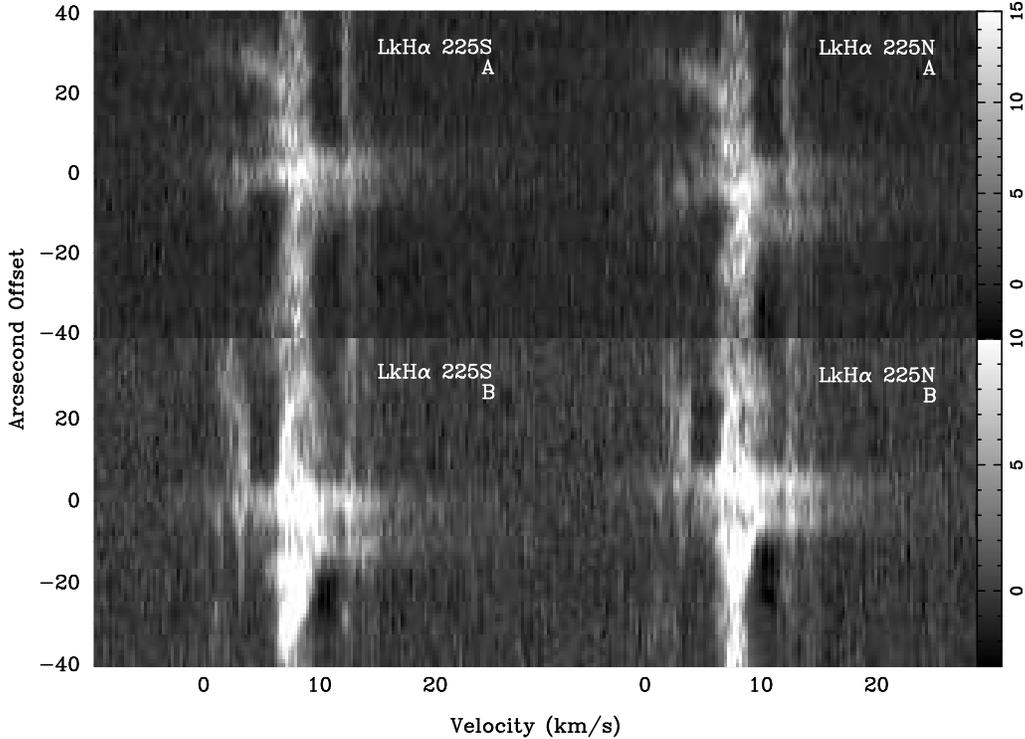}
\caption{Position-velocity diagrams taken through the two blue
  features marked A and B in Figure \ref{225co_mom0} are shown aligned
  through both the \lkha\ 225S and \lkha\ 225N sources. It is not
  possible to discriminate whether feature ``A'' arises from \lkha\
  225S or \lkha\ 225N. Feature B shows virtually the same profile
  through both sources.  Since this feature is aligned with the offset
  between the sources, discrimination is naturally difficult. There is
  still some evidence for the pole-on outflow of \lkha\ 225S (see
  Fig.\ \ref{pv_225s}) in slices through \lkha\ 225N. }
\label{225pv_bluefeatures}
\end{figure*}

To ascertain the physical origin of the ``ring'', inspection of
position-velocity ($P-V$) diagrams is critical.  Figure \ref{pv_225s}
shows the difference between two $P-V$ slices, one taken centered on
\lkha\ 225S and one taken at a position offset 24\arcsec\ to the
south, with the same orientation. Taking the difference of these two
$P-V$ diagrams crudely removes the large scale emission from three
separate molecular cloud components along the line of sight in Cygnus
(at roughly 3 \kms, 8 \kms\ and 13 \kms) and clearly reveals the
presence of emission at the 0\arcsec\ offset across both red and
blue-shifted velocities (from 0 to 20 \kms) with respect to the
source.  The $P-V$ diagram produces this feature no matter what the
orientation of the slice centered on \lkha\ 225S.  We interpret this
emission as arising from an outflow driven from \lkha\ 225S but
observed almost pole-on.  This is consistent with the $^{12}$CO
$J=2-1$ emission detected by \cite{pal95} in which they find strongly
overlapping blue and red lobes in the single detected outflow. The
blue-shifted emission is offset slightly to the east compared to the
red lobe, which is also consistent with the plane-of-sky emission
shown in Figure \ref{225co_mom0}, but is in the opposite sense to the
outflow lobes observed by \cite{pal95}.

Figures \ref{225co_mom0} and \ref{225co_channel} provide further
clarification in identifying the driving sources of the outflows with
components in the plane of the sky. In addition to the pole-on
outflow, a second outflow appears well-centered on \lkha\ 225, based
on the blue lobe of emission detected southeast of the two sources at
a position angle of $\sim 145^\circ$ (E of N).  Another blue peak is
detected to the northeast of the cluster of sources (at $\sim
45^\circ$), suggesting a third outflow is present in the system.
There is no evidence of coherence in the blue-shifted emission peaks
to suggest they are part of a larger, blue-shifted counterpart to the
red-shifted ``ring'', i.e., part of the same outflow.

Figure \ref{225pv_bluefeatures} shows the $ P-V$ diagrams across two
position angles suggested by the orientation of the blue-shifted
emission detected in the zeroth moment and channel maps (labeled ``A''
and ``B'' in Fig.\ \ref{225co_mom0}). Along these slices, there is
evidence for red-shifted components complementary to the blue-shifted
emission.  The red-shifted emission from these two potential outflows
appears juxtaposed on the zeroth moment map of Figure
\ref{225co_mom0}, lying coincident with the red-shifted gas from the
pole-on outflow and creating a ring structure which persists over a
range of velocities, although this emission appears significantly
closer to the driving sources than the blue lobes.

\begin{figure}
\vspace*{7cm}
\includegraphics{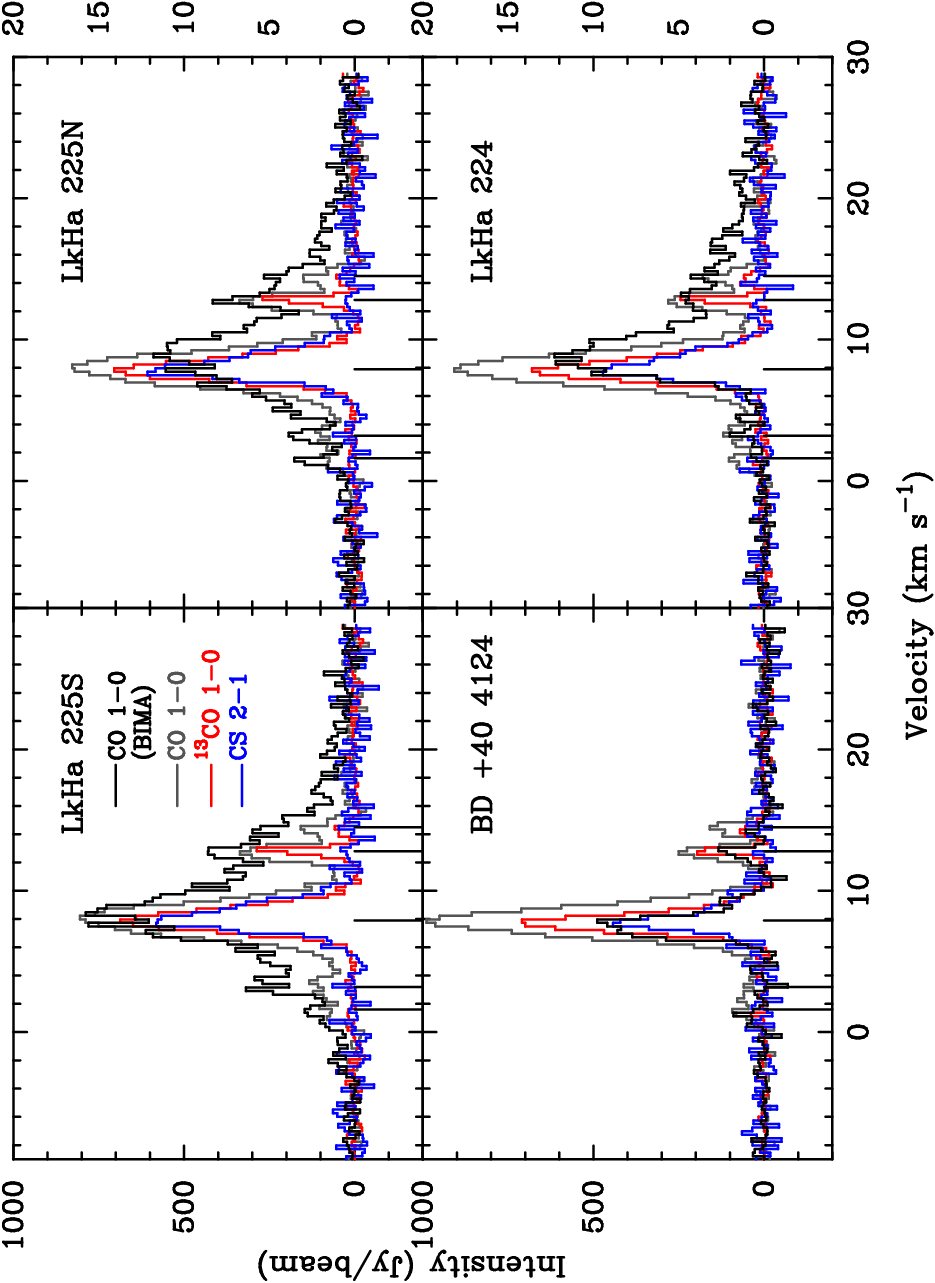}
\caption{Spectra toward individual sources in the \lkha\ 225S region
  from FCRAO and FCRAO+BIMA array data.  The spectral
  resolution is 0.254 \kms.  Grey, red and blue spectra represent \co, \tco\ and
  \cs\ observations from FCRAO which were regridded to the spatial and
  spectral scales of the FCRAO+BIMA \co\ map.  The intensity scale at
  left is for the single dish \co\ emission.  The \tco\ and \cs\
  emission have been multiplied by factors of two and twenty.  The
  black spectra are taken from the combined \co\ map. The intensity
  scale is shown at right.  Unlike the sources in the \lkha\ 198
  region, those in Cygnus do not exhibit significant self-absorption.
  \lkha\ 225S and \lkha\ 225N show the most extensive line wings.
  BD +40 4124 does not show any evidence of outflow while \lkha\ 224
  shows no high velocity blue-shifted emission, but extensive
  red-shifted line wings which could be associated with another source
  in the field. The lines below the plots indicate 1.6, 3.2, 7.9, 12.8
  and 14.5 \kms.  These peaks each represent individual clouds along
  the line of sight.}
\label{225_hires_spectra}
\end{figure}

\subsubsection{Spectra}

\begin{figure}
\vspace*{6cm}
\includegraphics{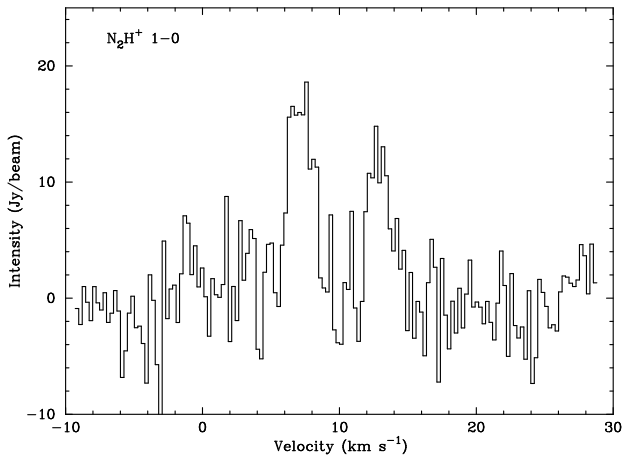}
\caption{Spectrum of \nthp\ emission toward \lkha\ 225S, indicating
  the presence of dense gas.}
\label{225S_nthp}
\end{figure}

Using FCRAO, we have detected \co, \tco, \cs\ and \nthp\ emission from
the \lkha\ 225S region.  The \co, \tco\ and \cs\ spectra all indicate
multiple emission peaks along the line of sight and are shown toward
each of the four sources in Figure \ref{225_hires_spectra}.  We also
show the combined \co\ spectra at high spatial and spectral
resolution.  We show an \nthp\ spectrum toward \lkha\ 225S in Figure
\ref{225S_nthp}. Strong evidence for outflowing gas is detected toward
\lkha\ 225S and \lkha\ 225N, but none is detected toward BD +40 4124
and only evidence for red-shifted emission is seen at the position of
\lkha\ 224.  We interpret the high velocity emission toward \lkha\ 224
as arising from the outflows generated by the \lkha\ 225 sources. Less
ambient CO appears to be present in this region compared to the
environment around \lkha\ 198, since no significant self-absorption is
observed toward any of the sources in Figure \ref{225_hires_spectra}.
Interpretation of these spectra is further complicated by the presence
of multiple clouds along the line of sight.  Peaks are observed at
five different velocities in the spectra (1.6, 3.2, 7.9, 12.8 and 14.5
\kms\ as identified by eye in Figure \ref{225_hires_spectra}). The
primary cloud, at 7.9 \kms, appears to contain all the HAeBe stars. A
second cloud, at 12.8 \kms, appears to contain significant amounts of
\co\ and \tco, but no \cs\ was detected at this velocity. The evidence
for clouds at the remaining velocities is based on \co\ only.

The resolution of FCRAO is such that we cannot be sure that all four
sources are associated with dense gas.  As the most embedded object,
\lkha\ 225S is the most likely source to be associated with dense
gas. However, BD +40 4124, which is the furthest removed from \lkha\
225S, lies only about 45\arcsec\ distant. This is only one FCRAO
beamwidth at the frequency of the \tco\ emission, and substantially
less than a beamwidth in \cs\ and \nthp. Higher resolution
observations can isolate the concentrations of these species and
enable kinematic study of the associated cores.

\subsubsection{2.6 mm Continuum}

No signficiant continuum was detected toward any of the sources in the
field containing \lkha\ 225S.  Our sensitivity is rather low, given
the limited bandwidth available for continuum observations.  We can
put an upper limit of 51 \mjybeam\ on the continuum flux on compact
sources in the region. Despite the strong CO and \nthp\ detections, we
see no flux to attribute to either dust continuum or free-free
emission.  \cite{dif97} detected 2.7 mm emission from \lkha\ 225S
using the Plateau de Bure interferometer with a resolution of
6.8\arcsec\ $\times$ 4.6\arcsec.  They detected a peak flux density of
23 \mjybeam\ and a total flux density of $24 \pm 6.7$ mJy.  This flux
would not have been detectable at a 2 $\sigma$ level with the
sensitivity ($\sigma = 17$ \mjybeam\ rms) in our map.

\subsection{Individual Sources}

Table \ref{sources_lkha225S} summarizes our findings regarding \co\ and
outflows in this region.  Self-absorption is not detected toward any
sources in this field. However, we do detect tracers of dense gas,
including \tco, \cs\ and \nthp\ in this cloud. We note also that our
interpretations of outflows from \lkha\ 225S and \lkha\ 225N are still
resolution-limited. Therefore, we discuss these sources together
below. 

\begin{deluxetable*}{llllcccc}
\tablecolumns{8} 
\tablewidth{0pc} 
\tablecaption{\lkha\ 225S \& Surrounding Objects}
\tablehead{\colhead{Source} & \colhead{Other ID} & 
\colhead{RA} & \colhead{DEC} & \colhead{Group$^a$} &
\colhead{Associated} & \colhead{\# Associated} & \colhead{PA} \\
& & \colhead{(J2000)} & \colhead{(J2000)} & & \colhead{CO} &
\colhead{Outflows} & \colhead{($^\circ$ E of N)} }
\startdata 
\lkha\ 225S & V1318S Cyg & 20:20:30.6 & +41:21:25.5 & ... & yes & 1-3$^b$ &
l.o.s.$^c$, 45\degr, 145\degr\ \\ 
\lkha\ 225N & V1318N Cyg & 20:20:30.6 & +41:21:30.5 & ... & maybe & 0-2$^b$
& 55\degr, 160\degr\ \\ 
BD +40 4124 & V1685 Cyg & 20:20:28.25 & +41:21:51.6 & I & yes & none & -- \\
\lkha\ 224 & V1686 Cyg & 20:20:29.26 & +41:21:28.6 & I & yes & none &
-- \\
\enddata 
\tablecomments{$^a$ Classification of \cite{hil92}. $^b$ The driving
  source of two outflows cannot be uniquely identified as \lkha\ 225S
  or \lkha\ 225N due to remaining resolution limitations. $^c$
  line-of-sight. The outflow is almost pole-on, with little component
  in the plane of the sky.}
\label{sources_lkha225S}
\end{deluxetable*}

\subsubsection{\lkha\ 225S and \lkha\ 225N}

\lkha\ 225S is found to have evidence for up to three associated
outflows. The most definitive connection is based on evidence for a
nearly pole-on outflow clearly seen in the $P-V$ diagram of Figure
\ref{pv_225s}.  Evidence for two additional outflows is seen in Figure
\ref{225co_mom0} in the form of two blue-shifted emission features.
The solid lines show the axes formed by the features, ``A'' and ``B'',
with \lkha\ 225S and both axes align with red-shifted features seen on
the other side of the potential driving source. However, \lkha\ 225S
is separated from \lkha\ 225N by only 5\arcsec, which is comparable to
the major axis of our synthesized beam (naturally-weighted). This
makes it difficult to associate decisively either of these two
plane-of-sky outflows with \lkha\ 225S.  Based on alignment, a strong
case can be made that \lkha\ 225N is driving one or both of these
outflows (see the dashed lines of Figure \ref{225co_mom0}).

We assert that the apparent ring of Figure \ref{225co_mom0} has
coherent structure, as it is dominated by the red-shifted emission of
the pole-on outflow with contributions from the other outflows
originating from the \lkha\ 225 sources, one oriented at approximately
45-55\degr\ E of N, the second at about 145-160\degr\ E of N and the
third almost pole-on toward us.  No evidence of the ring morphology is
observed in previous detections of high-velocity gas in the region
\citep[e.g., ][]{pal95} with resolution only a factor of three poorer
than our data.

As we noted above, the interpretation of the ring as the red-shifted
counterparts to the more definitive blue lobes on the plane of the sky
suggests that the red lobes have not expanded to the same spatial
extent. The recent work of \cite{loo06} suggests a potential
explanation. Their high resolution BIMA array observations of \cs\
show that the \cs\ traces out a dense filament of emission to the
northwest of the \lkha\ 225 sources.  The presence of denser gas in
this direction may be impeding the red-shifted component of the
outflow from extending into the cloud along these position angles.

\subsubsection{BD +40 4124}

BD +40 4124 is a Group I HAeBe star as classified by \cite{hil92}.  This
means that there is no remaining circumstellar emission which cannot
be accounted for by a disk.  \co, \tco\ and \cs\ are all detected
toward BD +40 4124 based on the spectra of Figure
\ref{225_hires_spectra}.  This emission is likely associated with the
ambient molecular clouds along the line of sight to Cygnus. The lack
of self-absorption indicates that relatively little gas remains around
this source, and there is no significant evidence of line wings in the
spectra.

\subsubsection{\lkha\ 224}

\lkha\ 224 is also classified as a Group I HAeBe star by \cite{hil92}.
The spectrum of \lkha\ 224 is dominated by the multiple peaks
associated with cloud emission along the line of sight. The combined \co\
spectrum does show evidence of a red-shifted line wing however.  No
outflow emission appears focussed on this source based on the channel
maps of Figure \ref{225co_channel}, and no blue-shifted emission is in
the vicinity of this source in the zeroth moment map of Figure
\ref{225co_mom0}.   Given its position relative to the blue-shifted
emission peaks, we surmise that the red-shifted emission observed
at the position of this source is part of one of the outflows
originating from the \lkha\ 225 sources.

\section{Conclusions}
\label{sum}

We find that the single bipolar outflow detected in the \lkha\ 198
region in single-dish maps is a combination of the red-shifted
emission of the \lkha\ 198 and \lkha\ 198-mm outflows and the
blue-shifted components of the outflows from \lkha\ 198-mm and V376
Cas.  In all, we find evidence for four outflows in the \lkha\ 198
region using high-resolution
\co\ observations.  Evidence for a cavity is observed in the \lkha\ 198
system along the same axis as the elliptical cavity detected in
optical and IR emission.  We see no evidence for circumstellar
material based on \co\ emission associated with \lkha\ 198-IR,
contrary to the results from near-IR polarimetry which associate dust
emission with that source \citep{per04} and observations of HH
objects, which identified \lkha\ 198-IR as a driving source
\citep{cor95}.  

Our ability to discriminate between detected \co\ outflows and measure
independent masses and momentum is limited by the single pointing of
the BIMA array maps (Figure \ref{198co_FCRAO_BIMA} clearly shows that
we have not mapped the outflows in their entirety) and spatial
resolution. The outflow emission is still entangled and the cloud
emission is also a contaminant. Ideally, mosaicked observations with
various array configurations (including very close baselines) will be
obtained to maximize the detection of the outflow emission, but minimize
the detection of emission from the ambient cloud.

The \co\ emission from \lkha\ 225S is equally complex.  In all, we
find evidence for up to three outflows in this region, all originating
from the \lkha\ 225 sources. Large-scale cloud emission dominates the
field, and multiple clouds along the line of sight contaminate the
spectra.  One is definitively associated with \lkha\ 225S and is
almost entirely along the line of sight.  The other outflows lie at
angles of 35\degr\ and 135\degr\ east of north projected onto the
plane of the sky.  We suggest that the ring of red-shifted emission is
in fact a juxtaposition of the red-shifted lobes of these two
outflows, where their expansion has been hindered due to a dense
molecular cloud lying to the northwest, with the cavity of the pole-on
outflow (a coherent structure).  As in the \lkha\ 198 region, higher
resolution, mosaicked maps would help disentangle the outflows from
the \lkha\ 225 sources.

Estimation of the mass in these outflows is not easy based on the data
presented here because they can only be resolved within the BIMA array
antenna field of view, whereas the outflows certainly extend to larger
spatial scales.  In several cases, emission from several outflows is
clearly combined, further complicating the independent assessments of
masses.

Finally, we find evidence of higher density gas in both regions, with
\tco\ and \cs\ detected toward all eight sources and \nthp\ detected
in the objects in Cygnus. Localizing the precise sources of this
emission requires higher resolution observations, since the separation
of sources is comparable or smaller than the FCRAO beam.

\acknowledgements

The authors would like to thank Mark Heyer for advice and observing
the FCRAO data for us.  We thank J. Di Francesco for useful
discussions and a careful reading of the manuscript.  BCM's research
was supported by an NSERC PDF and Berkeley NSF grant AST 02-28963.
MDP was supported by a NASA Michelson Fellowship.  The BIMA array was
operated with support from the National Science Foundation under
grants AST-02-28963 to UC Berkeley, AST-02-28953 to U.\ Illinois, and
AST-02-28974 to U.\ Maryland.  FCRAO was supported by NSF Grant AST
02-28993.  This work has been supported by the National Science
Foundation Science and Technology Center for Adaptive Optics, managed
by the University of California at Santa Cruz under cooperative
agreement No. AST - 9876783.

\end{document}